\documentclass[aps,prc,superscriptaddress,showpacs,nofootinbib,floatfix,twocolumn]{revtex4}

\usepackage{doi,graphicx,amssymb,amsmath}

\newcommand{\unit}[1]{\ensuremath{\, \mathrm{#1}}}
\newcommand{\etal}{\textit{et al. }}

\begin{document}

\title{Nuclear Pasta Formation}
\author{A. S. Schneider}\email{andschn@indiana.edu}
\author{C. J. Horowitz}\email{horowit@indiana.edu}
\author{J. Hughto}\email{jhughto@astro.indiana.edu}
\affiliation{Department of Physics and Nuclear Theory Center, Indiana University, Bloomington, IN 47405, USA}
\author{D. K. Berry}\email{dkberry@iu.edu}
\affiliation{University Information Technology Services, Indiana University, Bloomington, IN 47408, USA}
\date{\today}
\begin{abstract}
The formation of complex nonuniform phases of nuclear matter, known as nuclear pasta, is studied with molecular dynamics simulations containing 51\,200 nucleons.   A phenomenological nuclear interaction is used that reproduces the saturation binding energy and density of nuclear matter.
Systems are prepared at an initial density of $0.10\unit{fm}^{-3}$ and then the density is decreased by expanding the simulation volume at different rates to densities of $0.01\unit{fm}^{-3}$ or less.  
An originally uniform system of nuclear matter is observed to form spherical bubbles (``swiss cheese''), hollow tubes, flat plates (``lasagna''), thin rods (``spaghetti'') 
and, finally, nearly spherical nuclei with decreasing density.   
We explicitly observe nucleation mechanisms, with decreasing density, for these different pasta phase transitions.  
Topological quantities known as Minkowski functionals are obtained to characterize the pasta shapes.  
Different pasta shapes are observed depending on the expansion rate.  This indicates non equilibrium effects.  
We use this to determine the best ways to obtain lower energy states of the pasta system 
from MD simulations and to place constrains on the equilibration time of the system.
\end{abstract}


\maketitle

\section{Introduction}\label{sec:Intro}
During a supernova, the core of a massive star undergoes an extraordinary transformation, from $10^{55}$ separate nuclei to a single gigantic nucleus that forms the proto-neutron star.  
This transformation likely involves a series of nuclear pasta phase transitions that occur at densities somewhat below nuclear saturation density, $n_0\simeq0.16\unit{fm}^{-3}$, and involve a range of exotic nuclear shapes. 
Knowing how matter organizes itself as its density increases from $n\lesssim 0.1 n_0$ to $\sim n_0$ has been a long standing problem in nuclear physics \cite{Baym1971225,Williams1985844}.  
The description of nuclear matter at these subnuclear densities is relevant for a variety of problems such as determination of the structure and properties of neutron stars \cite{Baym1971225,Pethick19987,Watanabe2000455}, the equation of state of nuclear matter \cite{PhysRevLett.41.1623,Williams1985844} and neutrino transport in supernovae \cite{PhysRevC.69.045804,PhysRevC.70.065806}. 

It is well established that at low densities, $n\lesssim 0.1 n_0$, attractive short-range strong interactions correlate nucleons into (almost) spherical nuclei.
The size and shape of these nuclei are limited by Coulomb repulsion between protons and the surface energy of the system \cite{PhysRevLett.50.2066}.
Also well established is the fact that at high densities, $n\gtrsim n_0$, nuclear matter saturates and becomes uniform.  
However, in between these two limits the picture is much less clear and considerable effort has been devoted to determine the phase transitions of matter as it goes from one extreme to the other. 

Works using a liquid-drop model for the nuclei showed that a system with fixed temperature and proton fraction favors the formation of large nuclei, $Z>100$, as the density increases \cite{Baym1971225,PhysRevLett.41.1623}.
Lamb \etal \cite{PhysRevLett.41.1623} then showed that when the fraction of volume occupied by nuclei reached $\sim1/2$ matter would ``turn inside out'' and the system would then be composed of dense matter with bubbles of less-dense matter immersed in it.  
Later on, pioneering work by Ravenhall \etal \cite{PhysRevLett.50.2066} and Hashimoto \etal \cite{PTP.71.320} showed that, at densities just below nuclear saturation density, matter can organizes itself into other complex shapes besides spheres and spherical holes.  
This collection of shapes which includes rods, slabs and cylindrical holes is caused by \textit{frustration} of the system and is now referred to as \textit{nuclear pasta}.
The reason for frustration, the inability of a system to minimize all its fundamental interactions, is the competition between short-range nuclear attraction and long-range Coulomb repulsion \cite{Pethick19987}.

Many works based on a compressible liquid-drop model \cite{PhysRevLett.50.2066,PTP.71.320,Watanabe2000455,PhysRevC.87.028801} have explicit assumptions about the geometrical shapes of nuclear pasta.  Some even include more exotic phases such as gyroid and double-diamond morphologies \cite{PhysRevLett.103.132501}.
Approaches to the problem that do not explicitly assume any shape for the nuclear pasta have also been considered.  They include calculations based on the Thomas-Fermi approximation \cite{Williams1985844,Okamoto2012284,PhysRevC.65.045804,PhysRevC.85.055808}, Hartree-Fock methods \cite{PhysRevC.79.055801,1742-6596-426-1-012009,PhysRevLett.109.151101}, density-functional theory \cite{PhysRevC.72.015802}, relativistic mean field approximation \cite{PhysRevC.78.015802,PhysRevC.87.028801}, quantum molecular dynamics (QMD) \cite{PhysRevC.57.655,PhysRevC.66.012801,PhysRevC.68.035806,PhysRevC.69.055805,PhysRevC.77.035806,PhysRevLett.94.031101,PhysRevLett.103.121101} and semi-classical molecular dynamics (MD) \cite{PhysRevC.69.045804,PhysRevC.70.065806,PhysRevC.72.035801,PhysRevC.78.035806,PhysRevC.85.015807,PhysRevC.86.055805}.

All the works described in the paragraph above conclude that matter just below nuclear saturation density forms unusual structures with geometrical shapes that depend on temperature, proton fraction and density.  Some of these works, mainly the ones using a liquid-drop model and Thomas-Fermi approximation, use a Wigner-Seitz cell approximation to determine the periodicity of the pasta shapes \cite{PhysRevLett.50.2066,Watanabe2000455,PhysRevLett.103.132501,PhysRevC.72.015802,PhysRevC.85.055808,PhysRevC.87.028801} while other works use a unit cell to account for the periodicity of the system \cite{Williams1985844,PTP.71.320,PhysRevC.65.045804,PhysRevC.79.055801,1742-6596-426-1-012009,PhysRevLett.109.151101}.
Meanwhile, works based on QMD and MD methods use larger volumes and do not assume any periodicity in the pasta shapes besides the one imposed by periodic boundary conditions \cite{PhysRevC.57.655,PhysRevC.66.012801,PhysRevC.68.035806,PhysRevC.69.055805,PhysRevC.77.035806,PhysRevLett.94.031101,PhysRevLett.103.121101,PhysRevC.69.045804,PhysRevC.70.065806,PhysRevC.72.035801,PhysRevC.78.035806,PhysRevC.85.015807,PhysRevC.86.055805}.
Still, some QMD simulations were able to achieve some degree of periodicity within their simulation volumes \cite{PhysRevC.69.055805,PhysRevC.77.035806,PhysRevLett.94.031101,PhysRevLett.103.121101}.
Recently Okamoto \etal used the Thomas-Fermi approximation to calculate pasta shapes not limited to a single unit cell \cite{Okamoto2012284} and also obtained periodic configurations smaller than their simulation volume.

One reason for so many studies on the pasta phase is its relevance for properties of neutron stars and core collapse supernovae. 
For one, neutrino-pasta scattering helps determine the neutrino opacity in core collapse supernovae \cite{PhysRevC.69.045804,PhysRevC.70.065806}.  
This is because supernova neutrinos have wavelengths comparable to pasta sizes and can scatter coherently from the pasta.  
Also, electron pasta scattering is important for determining the shear viscosity, thermal conductivity \cite{PhysRevC.78.035806} and electrical conductivity.  
The electrical conductivity of the pasta may be relevant for the decay of neutron star magnetic fields \cite{arXiv:1304.6546}.

Possible hysteresis in pasta shapes with changes in density will contribute to the bulk viscosity.   
This could be important for the damping of collective r-mode oscillations of rapidly rotating neutron stars.  
The excitation spectrum is important for the pasta heat capacity, see also \cite{PhysRevC.72.035801}. 
Meanwhile, the shear modulus of the pasta is important for the speed of shear waves and crustal oscillation frequencies of neutron stars.  
The breaking strain of the pasta helps determine the maximum sized mountain that can be supported on a neutron star \cite{PhysRevLett.102.191102}.  
It may also be relevant for star quakes and crust breaking models of magnetar giant flares.  
In general the strength increases with increasing density and, therefore, the pasta is expected to make a significant contribution to the strength of the neutron star crust because of its high density.  

In the next sections, we will discuss some properties of the pasta phases at different densities obtained from molecular dynamics (MD) simulations.
We use the molecular dynamics formalism of Horowitz \etal \cite{PhysRevC.69.045804,PhysRevC.70.065806,PhysRevC.72.035801,PhysRevC.78.035806}.
However, while their main focus was transport properties of the pasta, our focus here is the study of the equilibration mechanisms and the topological structures formed by the pasta phases as they transition from low to high densities.  As the initial conditions are much simpler when the system is nearly uniform, we start our simulations at high densities near half the nuclear saturation density and then slowly expand the simulation volume to obtain lower densities.
We use Minkowski functionals to quantify how the pasta shapes change as a function of the expansion rate.  We believe this allows us to address some questions related to the equilibration time of the pasta and its transition from one phase to the other.  A couple of papers by Watanabe \etal have already explored the mechanisms of pasta phase transitions and their equilibration/transition times using QMD \cite{PhysRevLett.94.031101,PhysRevLett.103.121101}.  In these papers Watanabe \etal started their simulations from low density and compressed the system adiabatically or isothermally to determine the time scale of transitions between different pasta shapes.
As we will see the time scales they find using QMD are significantly faster than the ones we obtain with MD.

To study the pasta and its phase transitions it is helpful to provide some simple metrics that quantify their geometrical shapes.  
This can be achieved by making use of integral-geometric formulae \cite{lang01} often referred to as Minkowski functionals \cite{Michielsen2001461}. 
Minkowski functionals are a robust way to describe complex structures.
They were first used to describe the topology of nuclear pasta structures by Watanabe \etal in the context of quantum molecular dynamics \cite{PhysRevC.68.035806,PhysRevC.69.055805,PhysRevLett.94.031101}.
More recently Dorso \etal calculated the Minkowski functionals for pasta structures obtained from classical molecular dynamics \cite{PhysRevC.86.055805} while Schuetrumpf \etal did the same using a time-dependent Hartree-Fock approach  \cite{1742-6596-426-1-012009}.

In this paper we use MD to simulate nuclear matter for densities of $\sim0.010\unit{fm}^{-3}$ to $0.10\unit{fm}^{-3}$ for a proton fraction of $Y_p=0.40$ at a temperature of $T=1\unit{MeV}$.  This temperature and $Y_p$ are roughly comparable to those in the collapsing dense core of a supernova, before the matter is heated further by a shock wave.  We start Sec. \ref{sec:Formalism} reviewing the formalism used to describe nucleon-nucleon interactions, Sec. \ref{ssec:PastaForm}, and then move on to explain the methods used to obtain the topological properties of the pasta, Sec. \ref{ssec:Topology}.  In Sec. \ref{sec:Results} we present the main results of our simulations with Sec. \ref{ssec:Simulations} devoted to a description of our runs and Sec. \ref{ssec:Discussion} to a discussion of how the topology of the pasta changes as a function of density.
We conclude in Sec. \ref{sec:Conclusions}.

\section{Formalism}\label{sec:Formalism}

In Sec. \ref{ssec:PastaForm} we review our MD simulation formalism.
Section \ref{ssec:Topology} is devoted to the methods used to obtain the topology of the pasta while Sec. \ref{ssec:cluster} describes an algorithm to determine which nucleons belong to which nuclei. 

\subsection{Semiclassical nuclear pasta model}\label{ssec:PastaForm}

Here we briefly describe the formalism used in our MD simulations.  This is essentially the same as the one used by Horowitz \etal and others in previous works  \cite{PhysRevC.69.045804,PhysRevC.70.065806,PhysRevC.72.035801,PhysRevC.78.035806,PhysRevC.86.055805,PhysRevC.85.015807}.
The system is composed of neutrons, protons and electrons.
The electrons are assumed to be noninteracting and, thus, are described as a degenerate free Fermi gas.
Meanwhile, the nucleons are treated as point-like particles that interact via an ``elementary'' two-body interaction.  The interaction between any two nucleons $i$ and $j$ can be separated into nuclear, $v^n_{ij}$, and electromagnetic (Coulomb) $v^c_{ij}$, components; that is
\begin{equation}
 v_{ij}=v^n_{ij}+v^c_{ij}.
\end{equation}
The nuclear component of the interaction is
\begin{equation}
 v^n_{ij}=ae^{-r_{ij}^2/\Lambda}+\left[b+c\tau_z(i)\tau_z(j)\right]e^{-r_{ij}^2/2\Lambda}.
\end{equation}
Here, $r_{ij}=\vert\boldsymbol{r}_i-\boldsymbol{r}_j\vert$ is the distance between particles $i$ and $j$ 
and $\tau_z=+1$ $(-1)$ is the isospin projection of the particle if it is a proton (neutron).
The constants $a$, $b$, $c$ and $\Lambda$ describing the two-body potential are the same ones used in Ref. \cite{PhysRevC.69.045804}.
Their values are given in Table \ref{Tab:parameters} and were adjusted to approximately reproduce some bulk properties of pure neutron matter and symmetric nuclear matter as well as the binding energies of selected nuclei.

\begin{table}[h]
\caption{\label{Tab:parameters} Nuclear interaction parameters. The parameter $a$ defines the strength of the short-range repulsion between nucleons, $b$ and $c$ the strength of their intermediate-range attraction and $\Lambda$ the length scale of the nuclear potential.}
\begin{ruledtabular}
\begin{tabular}{*{4}{c}}
$a$ (MeV) &$b$ (MeV)&$c$ (MeV)&$\Lambda$ (fm$^{2}$) \\
\hline
  110     &  $-$26    &   24    &    1.25      \\
\end{tabular}
\end{ruledtabular}
\end{table}

The Coulomb component of the interaction is
\begin{equation}
 v^c_{ij}=\frac{\alpha}{r_{ij}}e^{-r_{ij}/\lambda}\tau_p(i)\tau_p(j),
\end{equation}
where $\alpha$ is the fine structure constant, $\tau_p\equiv(1+\tau_z)/2$ the nucleon charge and $\lambda$ is the screening length that results from the slight polarization of the background electron gas \cite{FW}.
The relativistic Thomas-Fermi screening length is given by
\begin{equation}
 \lambda=\frac{\pi^{1/2}}{2\alpha^{1/2}}\left(k_F\sqrt{k_F^2+m_e^2}\right)^{-1/2}
\label{eq.lambda}
\end{equation}
where $k_F=(3\pi^2n_e)^{1/3}$ is the Fermi momentum of the electrons with $n_e$ the electron density and $m_e$ the electron mass.
However, to be consistent with previous works, we fix $\lambda$ at a constant value $\lambda= 10\unit{fm}$.  This is somewhat smaller than Eq. \ref{eq.lambda} and allows us to decrease the size of our simulations without introducing large finite size effects.

All of our simulations have a fixed number of particles $N=51\,200$, a proton fraction of $Y_p=0.40$ and a temperature of $T=1\unit{MeV}$.  
We use periodic boundary conditions where a nucleon interacts only with the nearest periodic image of the other nucleons.  
We also use a cut-off radius of $8\lambda$ for the Coulomb part of the potential and $11.5\unit{fm}$ for the nuclear potential. 
Both potentials are assumed to be zero for distances larger than their cut-off radius.

Knowing the positions of each particle along with the inter-particle potentials allows us to calculate the total force on each nucleon.
The new particle positions and velocities are then obtained using a velocity-verlet algorithm \cite{PhysRev.159.98}.
After every time step $\Delta t$ we increase each side $l_i$ of the box ($i=x,y,z$) by $\Delta l=l_i(0)\dot\xi_i\Delta t$.
That is, the side $l_i$ of the box at a time $t$ is
\begin{equation}\label{eq:stretch}
 l_i(t)=l_{i}(0)\left(1+\dot{\xi}_it\right)
\end{equation}
where $l_i(0)$ is the initial length of the box and $\dot\xi_i$ is the expansion rate.
Particle positions are not adjusted artificially, rather, are allowed to respond dynamically to the changing simulation volume.
The velocities are unaffected by the stretching.

\subsection{Topology}\label{ssec:Topology}

A powerful and general method to quantify the topological structures present in a system is provided by the Minkowsky Functionals.
In $N$ dimensions there are $N+1$ Minkowski functionals which completely describe the morphological properties of an object.
In three dimensions the Minkowski functionals are quantities proportional to the volume $V$, surface area $A$, mean breadth $B$ and Euler characteristic $\chi$, see table \ref{Tab:MF}.

\begin{table}[h]
\caption{\label{Tab:MF} Minkowski functionals in three dimensions. Adapted from Ref. \cite{1742-6596-426-1-012009}.
$K$ is the domain where the functionals are evaluated while $\kappa_1$ and $\kappa_2$ are the principal curvatures on $\partial K$.} 
\begin{ruledtabular}
\begin{tabular}{ll}
$V$ & Volume \\
$A=\int_{\partial K} dA$ & Surface Area \\
$B=\int_{\partial K} \left(\kappa_1+\kappa_2\right)/4\pi\,dA$ & Mean Breadth\\
$\chi=\int_{\partial K} \left(\kappa_1\cdot\kappa_2\right)/4\pi\,dA$ & Euler Characteristic\\
\end{tabular}
\end{ruledtabular}
\end{table}

Though there are four Minkowski functionals two are sufficient to characterize the shapes of the pasta, the mean breadth $B$ and the Euler characteristic $\chi$.
The mean breadth $B$ is a measure of the average curvature of the structures that form the pasta and is proportional to the surface integral of the mean curvature $(\kappa_1+\kappa_2)/2$.
Here, $\kappa_1$ and $\kappa_2$ are the principal curvatures on the surface $\partial K$ that defines the pasta.
Recall that for a concave system the curvatures are negative while for convex systems the curvatures are positive.
The Euler characteristic $\chi$, though proportional to the surface integral of the Gaussian curvature $\left(\kappa_1\cdot\kappa_2\right)$, is related to the number of structures in the system. 
In the three dimensional case it can be shown to be given by the number of connected components plus the number of cavities minus the number of tunnels in the system \cite{Michielsen2001461}, that is
\begin{align}\label{eq:chi}
 \chi=&\quad\#(\text{connected components})\nonumber\\
      & + \#(\text{cavities}) - \#(\text{tunnels}).
\end{align}
The eight possible structures according to their curvatures are discussed in Tab. \ref{tab:phases}.
For images of these structures see Fig. 1 of Ref. \cite{1742-6596-426-1-012009}.

\begin{table}[h]
\caption{\label{tab:phases} Description of the possible pasta phases according to the values of the mean breadth $B$ and Euler characteristic $\chi$. 
Adapted from Refs. \cite{PhysRevC.86.055805,1742-6596-426-1-012009}.
Shapes include nearly-spherical nuclei (sph), three cylindrical or spaghetti phases classified according to their connectivity (rod-1,2, and 3), 
a flat sheet phase (slab), two hollow-tube or anti-spaghetti phases (rod-1 b and rod-2 b) and a nearly-spherical bubble phase (sph b).
}
\begin{ruledtabular}
\begin{tabular}{c|ccc}
           & $B<0$ &$B\sim0$& $B>0$ \\
\hline
$\chi>0$   & sph b &        & sph   \\
$\chi\sim0$&rod-1 b&  slab  & rod-1 \\
$\chi<0$   &rod-2 b&  rod-3 & rod-2 \\
\end{tabular}
\end{ruledtabular}
\end{table}

In our simulations we considered nucleons to be point particles. 
However, this makes the problem of calculating the four Minkowsky Functionals intractable as point particles do not define a surface.
To circumvent this we change our treatment of the particles from point-like to normal distributions centered at the position determined by the simulations.
Thus, the number density of a nucleon $i$ and, similarly, the charge density of a proton transforms as
\begin{equation}
 \delta(\boldsymbol{r}-\boldsymbol{r}_i)\rightarrow\frac{1}{\left(2\pi\sigma\right)^{3/2}}\exp\left(-\frac{\left(\boldsymbol{r}-\boldsymbol{r}_i\right)^2}{2\sigma^2}\right).
\end{equation}
where $\sigma$ is the standard deviation of the distribution.

With this in mind we follow the recipe laid out by Lang\etal  \cite{lang01} to obtain the four Minkowski functionals.
A short explanation of the method follows.

We start by dividing each side $l(t)$ of our cubic system into $n$ segments of edge length $\Delta$, that is, $l(t)=n\Delta$.
Now that the system has been divided into $n^3$ cubes with vertices at $\boldsymbol{r}=(i\Delta,j\Delta,k\Delta)$ with $i,j,k=0,...,n-1$
we fold a gaussian over each proton to determine the charge density $n_{ijk}$ at the vertex of each cube.
Our choice for charge density over nuclear density was made as there is more contrast in the former than the latter.
If the charge density $n_{ijk}$ is larger than a pre-defined threshold value $n_{\text{th}}$ that vertex is considered occupied and assigned a value $b_{ijk}=1$. 
Otherwise the vertex is considered unoccupied and $b_{ijk}=0$. Thus, we have a discrete binary image of our system.

Once all voxels, values of $b_{ijk}$, are determined we analyze its $2\times2\times2$-neighboorhood configuration.
The neighboorhood consists of eight voxels: 
$b_{ijk}$, $b_{ij+1k}$, $b_{ijk+1}$, $b_{ij+1k+1}$, $b_{i+1jk}$, $b_{i+1j+1k}$, $b_{i+1jk+1}$ and $b_{i+1j+1k+1}$.
Due to our choice of periodic boundary conditions if $i=n-1$ then $i+1=0$. The same is valid for $j$ and $k$.
This analysis consists of applying a filter to determine the grey-tone $g_{ijk}$ of each voxel.
The filter is chosen so that $g_{ijk}=\sum_{l,m,n=0}^1 2^{l+2m+4n}b_{i+l,j+m,k+n}$.
While the voxels $b_{ijk}$ define a 2-bit image as $b_{ijk}=0$ or $1$ the $g_{ijk}$ values define an 8-bit image as $g_{ijk}$ can assume any value from 0 to 255.
A simple histogram $h_l$ of the values of $g_{ijk}$ is enough to obtain the four Minkowski functionals.
For example, the occupied volume $V$ of the system is simply given by
\begin{equation}
 V=\Delta^3\sum_{l=0}^{127}h_{2l+1}
\end{equation}
while the Euler characteristic $\chi$ is given by
\begin{equation}
 \chi=\Delta^3\frac{\pi}{6}\sum_{l=0}^{255}\nu_l h_l
\end{equation}
for a suitable choice of vector $\nu_l$.
For a more in depth discussion and expressions for the surface area $A$ and the mean breadth $B$ see Ref. \cite{lang01}.

In our analysis we set the value of $\sigma$, which can be thought of as the nucleon radius, to 1.5\,fm.
Though this value for $\sigma$ is rather large for a nucleon radius it was chosen to ensure that at low densities, $n\lesssim0.010\unit{fm}^{-3}$, regions of very low charge density do not appear inside nucleon clusters.
Here, inside refers to regions bound by the threshold charge density $n_{\text{th}}$, discussed below.
The nucleon clusters at low densities, meanwhile, can be identified to be (almost) spherical nuclei, as shown later.
Regions of low charge density inside these clusters would be identified by our algorithm as holes inside a nucleus and we clearly should avoid that when determining the topology of the system.

\begin{figure}[h]
\centering
\includegraphics[width=0.5\textwidth]{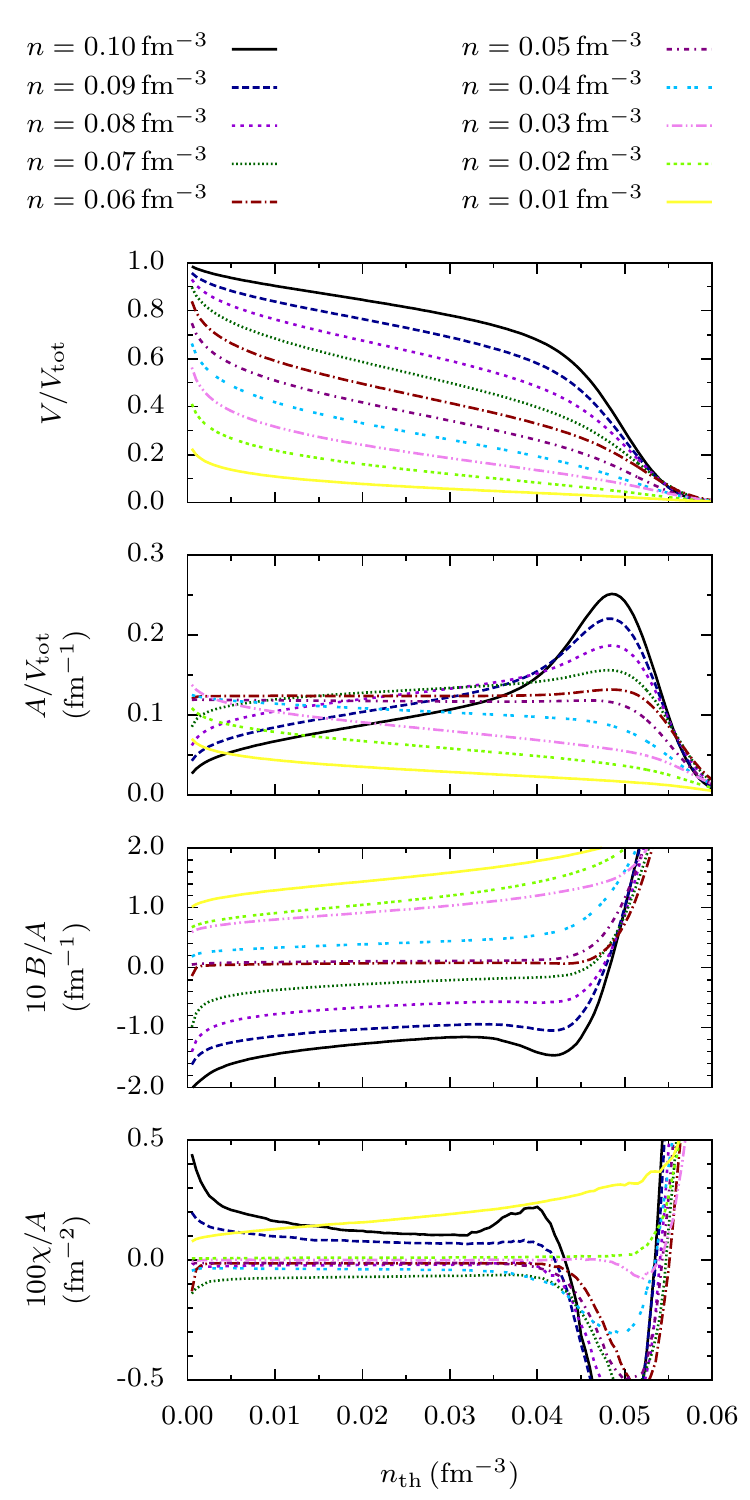}
\caption{\label{fig:MF} (Color on line) Normalized Minkowski functionals as a function of the threshold charge density $n_{\text{th}}$ for systems at different densities obtained stretching the pasta system from $0.10\unit{fm}^{-3}$ to $0.01\unit{fm}^{-3}$ at a rate of $\dot\xi=1.0\times10^{-7}\unit{c/fm}$. Lines are for the ten different densities $n$ indicated at the top.}
\end{figure}

Next we determine the threshold charge-density isosurface $n_{\text{th}}$ which will be used to to obtain the Minkowski functionals.
A comparison of normalized values for the four Minkowski functionals as a function of $n_{\text{th}}$ for one of our simulations is shown in Fig. \ref{fig:MF}.
The volume $V$ and surface area $A$ were normalized to the total volume of the system $V_{\text{tot}}$ while the mean breadth $B$ and the Euler characteristic were normalized to the total surface area $A$.
The procedure described by Michielsen and De Raedt in Ref.  \cite{Michielsen2001461} was shown to provide somewhat similar qualitative (though quantitatively different) results for our choice of $\sigma$.

In order to chose a value for $n_{\text{th}}$ to use we describe what sort of behavior we expect from the topology analysis of a low density system, $n\lesssim0.010\unit{fm}^{-3}$.
We want $n_{\text{th}}$ small enough that at these low densities, when nuclei can be clearly separated from each other, the charge isosurfaces $n=n_{\text{th}}$ determines a closed surface that contains all of the protons in a given cluster.
However, $n_{\text{th}}$ should also be large enough so that none of the protons in nearby clusters are contained within these isosurfaces.
This is similar to match at low densities the number of clusters obtained from the algorithm described in Sec. \ref{ssec:cluster} to the value of $\chi$ obtained from the analysis described in this section.
For these reasons we choose a threshold proton density of $n_{\text{th}}=0.030$\,fm$^{-3}$ as this value seems to satisfy both of the above requirements at low densities.
This value also makes the nuclei surfaces smooth enough that the Minkowski functionals converge for $\Delta\lesssim0.5\unit{fm}$ using the method outlined above.
We note that this procedure contrasts with the work of Watanabe \etal in Ref.  \cite{PhysRevC.68.035806} where they choose the isosurface $n=n_{\text{th}}$, 
which defines the shape of nuclear structures, to be based on nuclear density and as a function of the quantities $A$ and $V$.

\subsection{Cluster Algorithm}\label{ssec:cluster}

Now we explain our algorithm to determine the number of neutrons and protons in each cluster formed in the system.
The algorithm is very similar to the ``Minimum Spanning Tree'' (MST) described by Dorso \etal in  Ref. \cite{PhysRevC.86.055805} and Horowitz \etal in Ref. \cite{PhysRevC.70.065806}

\begin{figure}[h]
\centering
\includegraphics[width=0.5\textwidth]{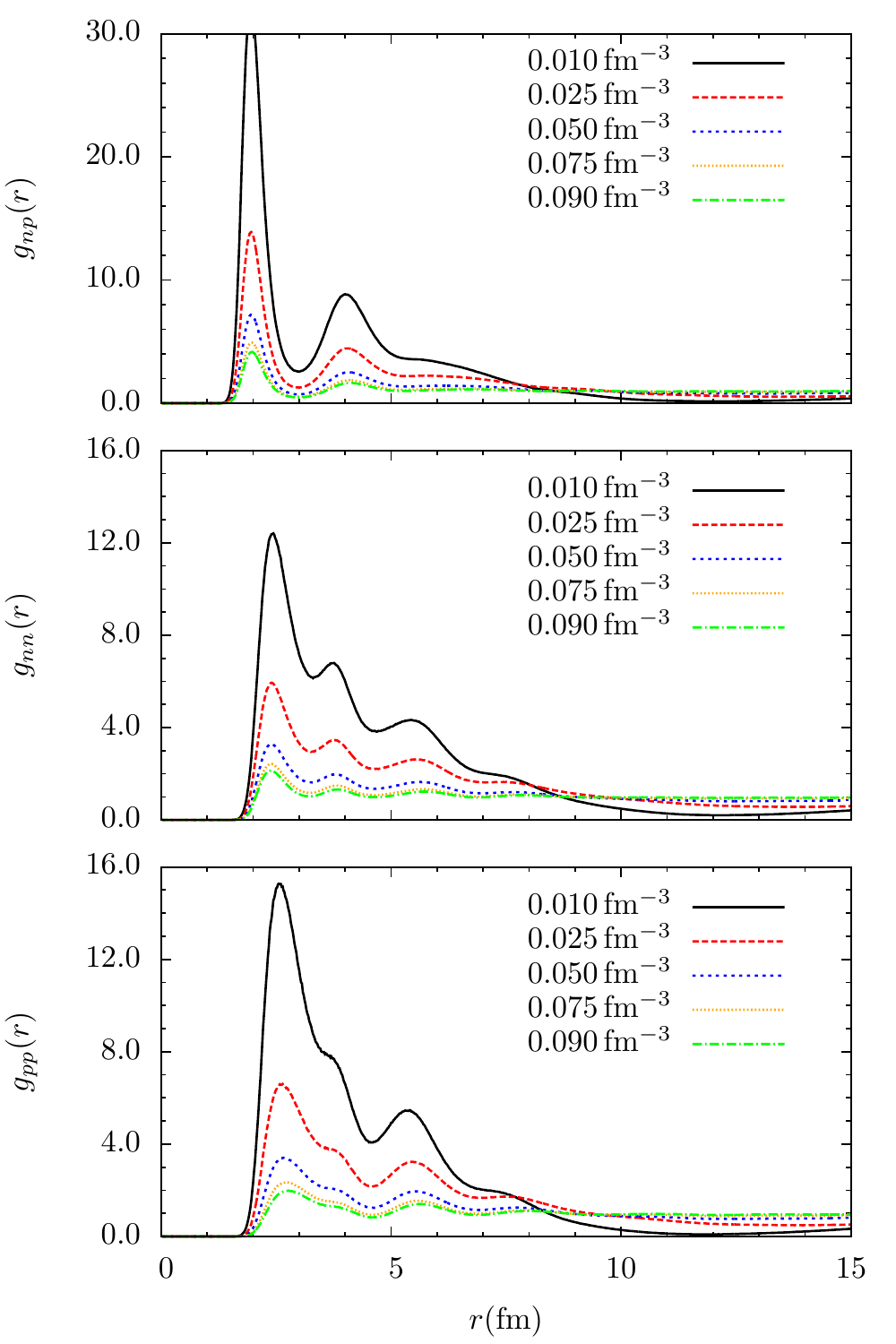}
\caption{\label{fig:gr} (Color on line) Nucleon pair-pair auto-correlation function as a function of pair distance for our five constant-density runs. See Sec. \ref{ssec:Simulations} for description of the five runs. The top plot shows neutron-proton correlations, the center plot shows neutron-neutron correlations while the bottom one shows proton-proton correlations.}
\end{figure}

Our algorithm starts by looking for correlations between the positions of protons.
A proton $i$ is said to be part of a cluster $C$ if it is within a cut-off distance $r_{pp}$ of at least one proton $j$ that is part of $C$.
That is, if nucleons $i$ and $j$ are protons then $i\in C$ if and only if $\exists j\in C$ such that $\vert\boldsymbol{r}_i-\boldsymbol{r}_j\vert\le r_{pp}$. 
We take $r_{pp}=4.5\unit{fm}$ as this is is very close to the first non-zero minimum of the proton-proton correlation function $g_{pp}(r)$.
We note that the position of this minimum does not appear to depend on the density of the system as can be seen on the bottom plot of Fig. \ref{fig:gr}.

After separating the protons into clusters we count the neutrons in each of the clusters.
We say a neutron is part of cluster $C$ if it is within a distance $r_{np}$ of at least one proton $j$ that is part of $C$.
That is, if $i$ is a neutron and $j$ is a proton then $i\in C$ if and only if $\exists j\in C$ and $\vert\boldsymbol{r}_i-\boldsymbol{r}_j\vert\le r_{np}$.
We take the cut-off distance $r_{np}=3\unit{fm}$, approximately the first non-zero minimum of the neutron-proton correlation function $g_{np}(r)$ as seen on the top plot of Fig. \ref{fig:gr}.
Neutrons that are not part of any cluster are considered free neutrons.

For completeness we also plot the neutron-neutron correlation function $g_{nn}(r)$ on the center of Fig. \ref{fig:gr}.

\section{Results}\label{sec:Results}

This section is devoted to the results of our simulations.
In Sec. \ref{ssec:Simulations} we detail the initial conditions and aspects of the different runs performed using the MD formalism.
Meanwhile, in Sec. \ref{ssec:Discussion} we describe the differences in pasta shapes at densities of 0.010, 0.025, 0.050, 0.075 and 0.090 $\unit{fm}^{-3}$ according to the type of simulation from which they were obtained.
We finish the section with a comparison of the Minkowski functionals as a function of density followed by a discussion of the equilibration time of the pasta. 

\subsection{Simulations}\label{ssec:Simulations}

Every simulation described here has 51\,200 particles, a proton fraction $Y_p=0.40$ and is isothermally evolved at a temperature of $1\unit{MeV}$.
Two types of simulation were performed in this work, stretching or expansion runs and constant-density runs.

For all of our expansion runs we set an initial density of $0.10\unit{fm}^{-3}$.
This makes our initial simulation volume a cube $80\unit{fm}$ on a side.
Each particle is given an initial random position inside the box and a velocity randomly selected from a Boltzmann distribution such that its temperature is $1\unit{MeV}$.
The system is evolved at constant density for a total time of $10\,000\unit{fm/c}$ using a time step of $\Delta t=1\unit{fm/c}$.
The temperature is kept constant by rescaling the velocities of nucleons every one hundred time steps.
Though the topological characteristics of these short equilibration runs do not change much after $10\,000\unit{fm/c}$ compared to much longer runs, $500\,000\unit{fm/c}$, it is not clear whether the system has equilibrated.
Further simulations, possibly starting at nuclear saturation density or above and slowly expanded to $0.10\unit{fm}^{-3}$, are needed in order to study the equilibrium configuration of the system. 
This will be done in future works. 
After the short equilibration time the final configuration is evolved using three different stretching rates, $\dot\xi=10^{-5}$, $10^{-6}$ and $10^{-7}\unit{c/fm}$, until it reaches a density of about $0.01\unit{fm}^{-3}$.
An output file with the positions and velocities of all particles in the system is written every $1\,000\unit{fm/c}$ and its topological characteristics obtained.   
An expansion run without the Coulomb interaction and a stretching rate of $\dot\xi=1.0\times10^{-6}\unit{c/fm}$ was also performed following the same procedure as the runs described above.

Five constant-density runs were performed at fixed densities of 0.010, 0.025, 0.050, 0.075 and 0.090 $\unit{fm}^{-3}$.
Initially we assign each particle a random position inside a cubic box with sides $l=\sqrt[3]{N/n}$.
Each particle is also given an initial velocity randomly selected from a Boltzmann distribution such that the system has a temperature close to $1\unit{MeV}$.  
The system is evolved at constant density for a total time of $500\,000\unit{fm/c}$ using a time step of $1\unit{fm/c}$.  
The velocities of the nucleons are rescaled every one hundred time steps to keep a $1\unit{MeV}$ temperature.  
The topological characteristics of the system are obtained every $1\,000\unit{fm/c}$ for the last $100\,000\unit{fm/c}$ of the run and are observed to not change significantly during that time.

With the exception of our slowest expansion run, $\dot\xi=1.0\times10^{-7}\unit{c/fm}$, all of our simulations were performed on the BigRed supercomputer at Indiana University using typically 128 cores for a few days to about a week.
The expansion run with $\dot\xi=1.0\times10^{-7}\unit{c/fm}$ was performed on the Kraken supercomputer \cite{Kraken} using about 1152 cores for a total of about 150 hours.

\subsection{Discussion}\label{ssec:Discussion}

\begin{figure*}[h]
\centering
\includegraphics[trim = 0.2in 0.5in 0.2in 0.5in, clip, width=1\textwidth]{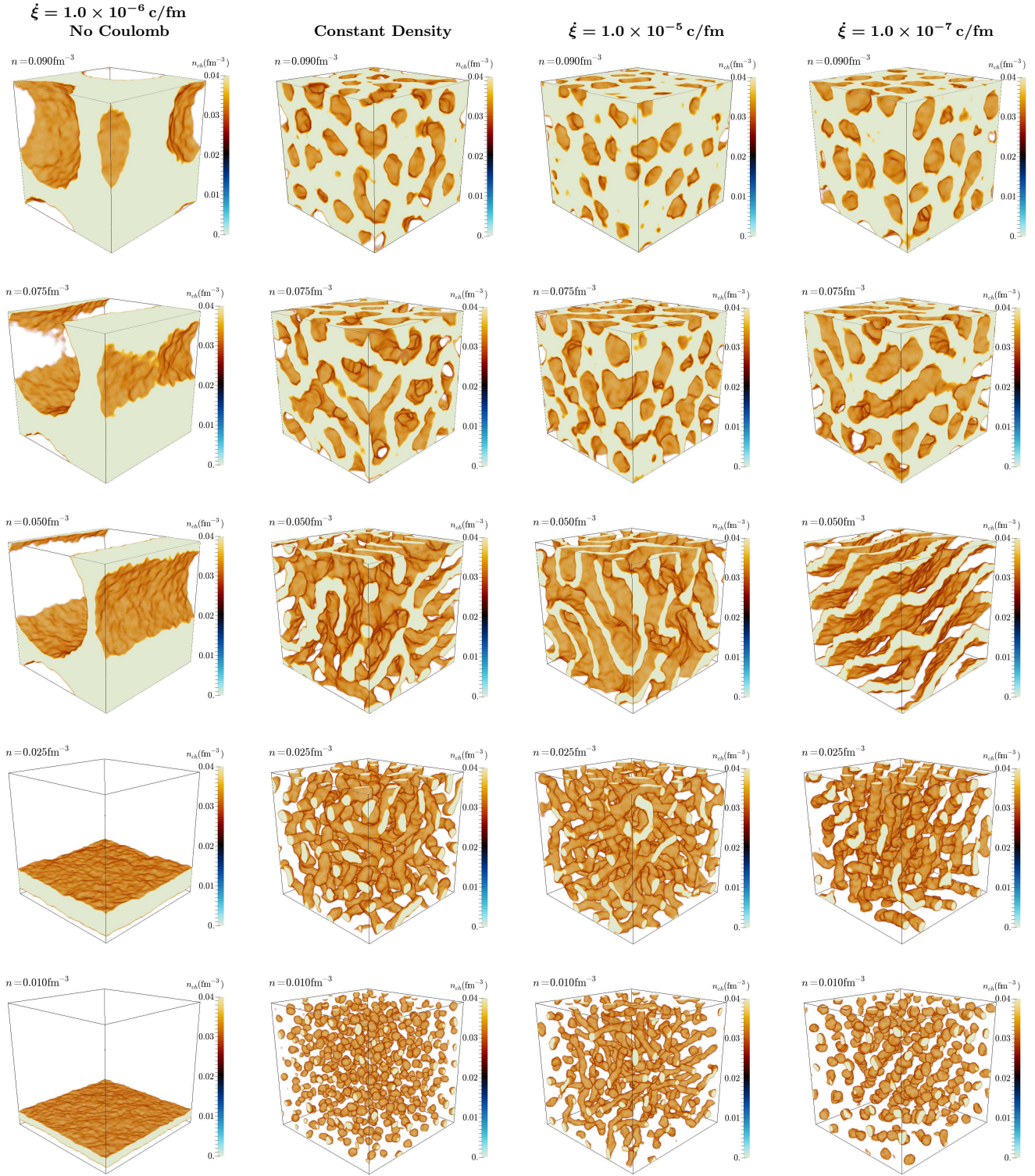}
\caption{\label{fig:pasta_phases} (Color on line) Side by side comparison of charge density isosurfaces of configurations with densities of 0.010, 0.025, 0.050, 0.075 and 0.090 $\unit{fm}^{-3}$ obtained from four different runs. On the left side we show configurations that were obtained neglecting Coulomb interactions.  Next we show constant density runs, and then configurations with the fastest of our stretching rates, $\dot\xi=1.0\times10^{-5}\unit{fm/c}$. Finally, to the right we show configurations with the slowest of our stretching rates, $\dot\xi=1.0\times10^{-7}\unit{fm/c}$. This figure was generated using ParaView \cite{Paraview}.}
\end{figure*}

In Fig. \ref{fig:pasta_phases} we compare the isosurfaces of charge density for four different runs at densities of 0.010, 0.025, 0.050, 0.075 and 0.090 $\unit{fm}^{-3}$.
For the constant-density run we show the isosurfaces obtained for the configuration at $500\,000\unit{fm/c}$. 
For the two stretching runs including Coulomb potential, $\dot\xi=1.0\times10^{-5}\unit{c/fm}$ and $\dot\xi=1.0\times10^{-7}\unit{c/fm}$, and for the run that does not include the Coulomb potential, $\dot\xi=1.0\times10^{-6}\unit{c/fm}$, we show the configuration closest to the desired density.
Values for the normalized Minkowski functionals $B/A$ and $\chi/A$ are listed in Tab. \ref{tab:ba} and Tab. \ref{tab:ca}, respectively.
These values can be used to describe the predominant phase of each configuration as discussed in Tab. \ref{tab:phases}.

We now describe the differences between the four systems shown in Fig. \ref{fig:pasta_phases} for each of the five densities mentioned.
This is intended to enlighten the reader about how the equilibration time affects our simulations.  
We will then discuss the evolution of the Minkowski functionals as a function of density and stretch rate.

\begin{table}[h]
\caption{\label{tab:ba} Values for the normalized mean breadth $B/A$ for our simulations at densities of 0.010, 0.025, 0.050, 0.075 and 0.090 $\unit{fm}^{-3}$.} 
\begin{ruledtabular}
\begin{tabular}{c|*{4}{c}}
$n$             & \multicolumn{4}{c}{$B/A$ (fm$^{-1}$)}                         \\
({fm}$^{-3}$)   & No Coulomb & Constant & $\dot\xi=10^{-5}$ & $\dot\xi=10^{-7}$ \\
\hline
0.090 & $-1.5\times10^{-2}$ & $-8.0\times10^{-2}$ & $-1.0\times10^{-1}$ & $-9.6\times10^{-2}$ \\
0.075 & $-4.3\times10^{-3}$ & $-3.4\times10^{-2}$ & $-6.8\times10^{-2}$ & $-4.4\times10^{-2}$ \\
0.050 & $-2.6\times10^{-3}$ & $+4.3\times10^{-2}$ & $+1.2\times10^{-2}$ & $+1.1\times10^{-2}$ \\
0.025 & $+5.9\times10^{-3}$ & $+1.2\times10^{-1}$ & $+1.0\times10^{-1}$ & $+1.1\times10^{-1}$ \\
0.010 & $-4.6\times10^{-3}$ & $+2.1\times10^{-1}$ & $+1.5\times10^{-1}$ & $+1.6\times10^{-1}$ \\
\end{tabular}
\end{ruledtabular}
\end{table}

\begin{table}[h]
\caption{\label{tab:ca} Values for the normalized Euler characteristic $\chi/A$ for our simulations at densities of 0.010, 0.025, 0.050, 0.075 and 0.090 $\unit{fm}^{-3}$.} 
\begin{ruledtabular}
\begin{tabular}{c|*{4}{c}}
$n$             & \multicolumn{4}{c}{$B/A$ (fm$^{-1}$)}                         \\
({fm}$^{-3}$)   & No Coulomb & Constant & $\dot\xi=10^{-5}$ & $\dot\xi=10^{-7}$ \\
\hline
0.090 & $+2.5\times10^{-5}$ & $+1.5\times10^{-4}$ & $+7.5\times10^{-4}$ & $+6.8\times10^{-4}$ \\
0.075 & $-2.2\times10^{-5}$ & $-7.4\times10^{-4}$ & $-2.9\times10^{-5}$ & $-4.5\times10^{-4}$ \\
0.050 & $-1.5\times10^{-5}$ & $-1.1\times10^{-3}$ & $-7.6\times10^{-4}$ & $-1.2\times10^{-4}$ \\
0.025 & $+2.0\times10^{-10}$& $+4.3\times10^{-4}$ & $-4.6\times10^{-4}$ & $+5.0\times10^{-5}$ \\
0.010 & $+1.5\times10^{-12}$& $+3.5\times10^{-3}$ & $+9.0\times10^{-4}$ & $+1.9\times10^{-3}$ \\
\end{tabular}
\end{ruledtabular}
\end{table}

\subsubsection{Systems at $n=0.090\unit{fm}^{-3}$}\label{ssec:90}

At a density of $0.090\unit{fm}^{-3}$ the four systems being compared have a negative value for $B/A$ and a positive value for $\chi/A$.
This is typical of systems formed mostly by spherical bubbles, see Tab. \ref{tab:phases}, and it happens because over most of the isosurface defined by $n_{\text{th}}$ both principal curvatures are negative, that is, $\kappa_1<0$ and $\kappa_2<0$.
Another thing to notice is how different the simulation that does not include the Coulomb potential looks from the simulations that include Coulomb potential, see first line of Fig. \ref{fig:pasta_phases}.
While the runs that include the Coulomb potential are mostly uniform with several spherical and cylindrical bubbles, the system without Coulomb is formed by two phases: one uniform matter phase and one large spherical bubble.
This differences can also be noted in the much smaller absolute values for $B/A$ and $\chi/A$ for that system when compared to the others.

Focusing on the simulations that include the Coulomb potential, we note that the one simulation ran at constant density has smaller absolute values for $B/A$ and $\chi/A$ than the ones obtained from stretching the box, see Tabs. \ref{tab:ba} and \ref{tab:ca}.
This is explained by the fact that the bubbles in this system are more elongated than the ones in the systems obtained from stretching the box from a higher density.
Also, the simulation ran at a constant density contains some hollow-tubes, cylindrical holes that stretch over the whole length of the box, while the others do not.
Comparing just the two systems obtained from stretching we observe that the slower the stretch rate the smaller the absolute values for $B/A$ and $\chi/A$. 
This is because the system stretched at a rate of $\dot\xi=1.0\times10^{-7}\unit{c/fm}$ has had more time to equilibrate and, thus, some of the spherical bubbles that existed in that system at a higher density had enough time to merge and form more elongated bubbles and even some tunnels.

\subsubsection{Systems at $n=0.075\unit{fm}^{-3}$}\label{ssec:75}

As the density decreases to $0.075\unit{fm}^{-3}$ the spherical bubbles merge to form tunnels and the value of the Euler characteristic $\chi$ in the four systems change from positive to negative.
Again the simulation without Coulomb potential has only two phases and absolute values for both average curvatures much smaller than the other simulations.
As for the constant density simulation that included the Coulomb potential, the large value of $\vert\chi/A\vert$ for the constant density run implies that system is mainly formed of interconnected connected tunnels, rod-2 b phase in Tab. \ref{tab:phases}.

Meanwhile, the configuration obtained from stretching the box quickly, $\dot\xi=1.0\times10^{-5}\unit{c/fm}$, could be identified looking up the average curvatures and Tab. \ref{tab:phases} as a system of hollow tubes, rod-1 b phase.
However, a glance at Fig. \ref{fig:pasta_phases} shows that is not the case. 
The system structure, besides hollow tubes, $\chi/A<0$, contains spherical bubbles, $\chi/A>0$ and is, thus, a mixture of the sph and rod-2 b structures.
As both of these phases contribute similarly to the Euler characteristic $\chi/A$ at this density a small value for this quantity is observed, see Eq. \eqref{eq:chi}.

The simulation that was stretched slowly, $\dot\xi=1.0\times10^{-7}\unit{c/fm}$, is mainly formed of tunnels that stretch over the whole length of the box.
Like the constant density run it can also be identified as a rod-2 phase, despite their different values for the average curvatures.
In this case the smaller absolute value for $\chi/A$ implies a lower interconnectivity between the tunnels.

\subsubsection{Systems at $n=0.050\unit{fm}^{-3}$}\label{ssec:50}

At $0.050\unit{fm}^{-3}$ the run without Coulomb potential exhibits the same behavior it does at $0.075\unit{fm}^{-3}$.
Meanwhile, all simulations with Coulomb interactions change from being mostly concave to mostly convex as show by the change in sign for the values of $B/A$.
The slowly stretched simulation, $1.0\times10^{-7}\unit{c/fm}$, has formed sheets that are almost parallel to each other, ``lasagna phase'', with some connectivity between the slabs.
Thus, the very small value of $\vert\chi/A\vert$ when compared to the fast stretched or constant density simulations. 
The system stretched at a rate of $1.0\times10^{-5}\unit{c/fm}$ also exhibits plane-like structures.
However, since the planes in this system have more connectivity amongst themselves than the slowly stretched system, the value of $\vert\chi/A\vert$ is much larger.
The constant-density run, though, has much larger values for $B/A$ and $\vert\chi/A\vert$ than the stretched systems.
The reason is that, unlike the other two systems, this one does not have plane-like structures and is formed of elongated nuclei connected to each other, rod-3 phase.
This behavior that produces slab or rod-3 phases depending on the initial condition has already been observed by Schuetrumpf \etal in Ref. \cite{1742-6596-426-1-012009}.
As they observed, in these two phases the liquid and solid phases have the structures that are symmetrical complements of each other.

\subsubsection{Systems at $n=0.025\unit{fm}^{-3}$}\label{ssec:25}

At even lower densities, $n=0.025\unit{fm}^{-3}$, the simulation ran without Coulomb potential has transitioned to a single liquid phase with nearly flat surfaces that have little curvature. 
Therefore, both the mean and Gaussian average curvatures $B/A$ and $\chi/A$ are very close to zero.
The configurations obtained for the other three simulations have similar values for the average mean breadth $B/A$.
This is because the systems with Coulomb interactions are formed of elongated nuclei, known as the ``spaghetti phase'', of about the same thickness.  
However their values for $\chi/A$ differ by large amounts. The reason for this is discussed below. 

The configuration obtained by quickly stretching the box has a large negative value for $\chi/A$.  
This value can be explained by a system made of one large nucleus that splits and reconnects several times, rod-2 phase.
Because of the many splits and reconnections the value of $\chi$ is negative and large\footnote{To understand this better we recall that an orientable surface $R$ has an Euler characteristic given by $\chi(R)=2-2g(R)$. 
Here $g(R)$ is the genus of that surface, that is, the number of ``handles'' in the surface.  
Thus, while a nucleus in empty space with  $g(R)=0$, or a cavity in uniform matter also with $g(R)=0$, have $\chi=2$, a torus with $g(R)=1$, has $\chi=0$ and a double-torus or ``pretzel'' with $g(R)=2$, has $\chi=-2$.}.
According to our clustering algorithm this system is formed of a nucleus with $A\sim47\,000$ and about 15 smaller nuclei with mass number in the range of $A=2$ to $A\sim2\,000$.
On the other hand, the simulation ran at constant density has a large positive value for $\chi/A$ as it is formed of several elongated nuclei with little to no splits and reconnections.  
From our clustering algorithm we see that this system is formed of 47 nuclei with mass number $A$ in the range of $A\sim150$ to $A\sim3\,200$.
In between these two extremes is the system obtained from slowly stretching the box which has a value for $\chi/A$ close to zero.  
Because of periodic boundary conditions, this system is formed by only three large spaghetti-like nuclei with $A\sim7\,000,\,17\,000$ and $27\,000$.
Since these nuclei do not curve or split/reconnect and are for the most part parallel to each other one of its principal curvatures will always be close to zero.  
Thus, the small value for $\chi/A$ for the slowly stretched system when compared to the other two systems.

\subsubsection{Systems at $n=0.010\unit{fm}^{-3}$}\label{ssec:10}

\begin{figure}[h]
\centering
\includegraphics[width=0.5\textwidth]{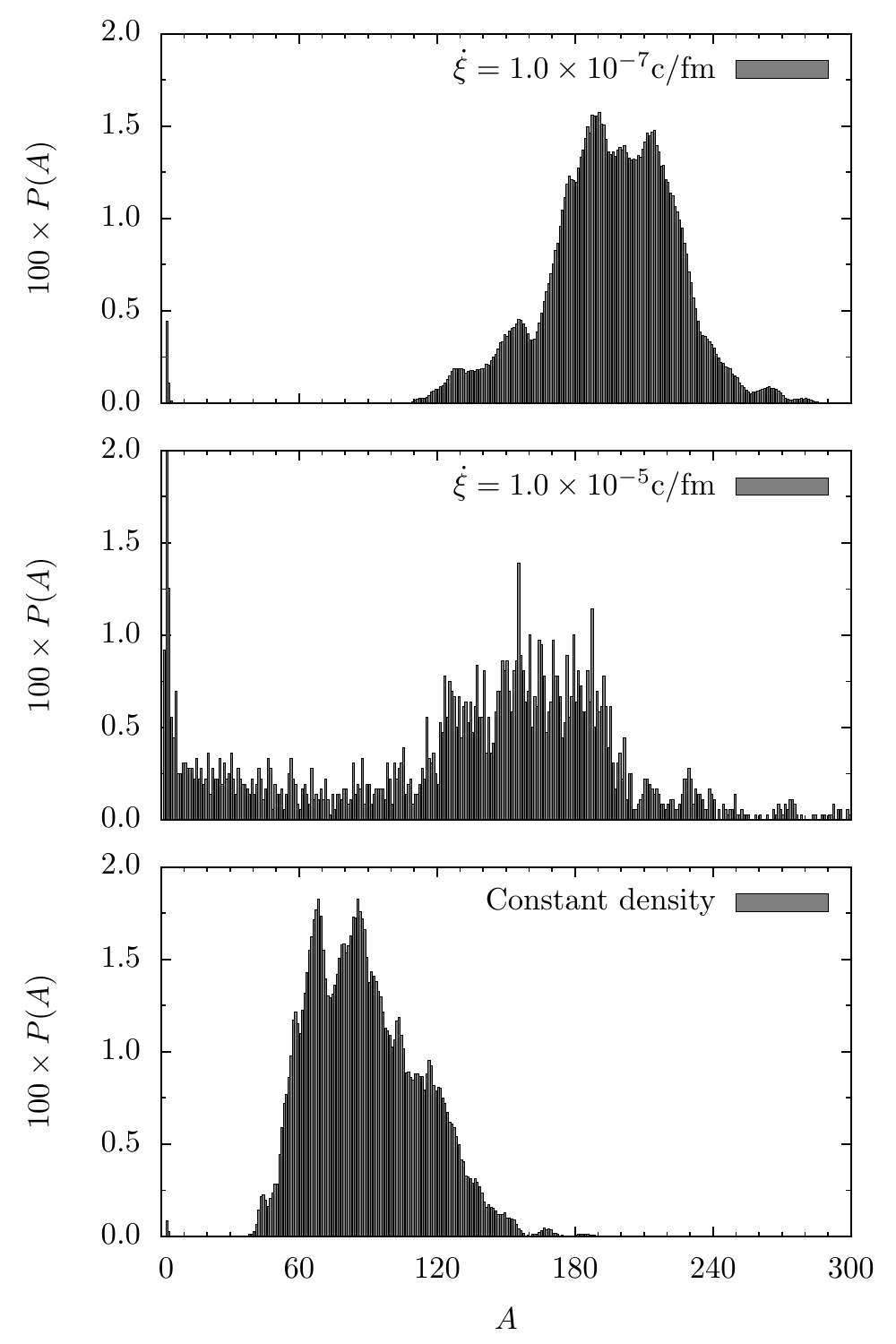}
\caption{\label{fig:hist} Probability of finding a nuclus with mass number $A$ at a density of about $0.010\unit{fm}^{-3}$ for three different runs: one at a constant-density (bottom), one obtained from stretching the box at $\dot\xi=1.0\times10^{-5}\unit{c/fm}$ (center) and one from stretching the box at $\dot\xi=1.0\times10^{-7}\unit{c/fm}$ (top).}
\end{figure}

At a density of $0.010\unit{fm}^{-3}$ both the system obtained from slowly stretching, $\dot\xi=1.0\times10^{-7}\unit{c/fm}$, and the system obtained from constant-density equilibration exhibit only (almost) spherical nuclei.  This is known as the ``gnocchi phase''.
From the top and bottom plots of Fig. \ref{fig:hist} one can tell that there is a difference in the size of the spherical nuclei formed.
While the slowly stretched system has spherical nuclei mostly in the range of $A\sim120$ to $300$, the constant-density run produces smaller nuclei, $A\sim40$ to $150$.
We note that there are barely any free neutrons in these systems and, therefore, the mass number and charge are related by $A\sim5Z/2$.
Meanwhile, the system stretched at a fast rate $\dot\xi=1.0\times10^{-5}\unit{c/fm}$ exhibits spherical nuclei of several sizes as well as elongated nuclei.
The center plot in Fig. \ref{fig:hist} shows that this system exhibits several very large nuclei, $A\gtrsim300$, as well as several very small ones, $A\lesssim20$. 
These difference are discussed below.   Finally for the simulation without Coulomb interactions, the single liquid phase now occupies a smaller fraction of the total simulation volume then at a density of $0.025$ fm$^{-3}$. 

\begin{figure}[h]
\centering
\includegraphics[width=0.5\textwidth]{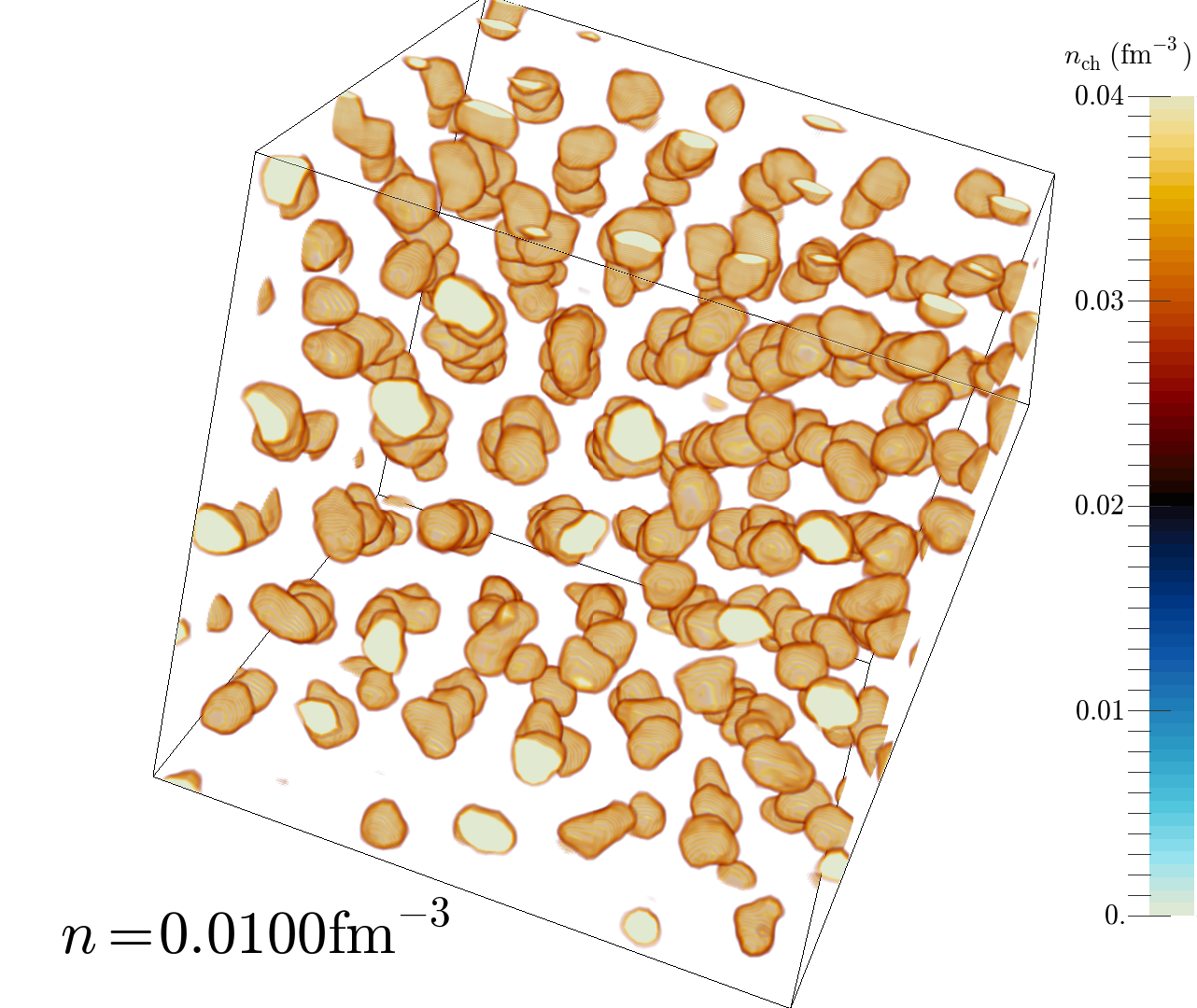}
\caption{\label{fig:bcc} System at a density of $0.010\unit{fm}^{-3}$ obtained from stretching the box at a rate of $\dot\xi=1.0\times10^{-7}\unit{c/fm}$ from $0.10\unit{fm}^{-3}$. 
The system is shown at an angle that makes it easier to see that the nuclei form some type of lattice. This figure was generated using ParaView \cite{Paraview}.}
\end{figure}

As mentioned, the system obtained from stretching the box at a slow rate has very few small nuclei $A\lesssim10$ while most have mass number in the range $A\sim120$ to $300$.
Also, the nuclei form a solid-like structure, possibly a $bcc$ lattice. One of the planes of the lattice can be seen clearly in Fig. \ref{fig:bcc}.
A reason to believe the lattice has a $bcc$ structure is its temperature $T$, average charge $\langle{Z}\rangle$ and the ratio $\kappa=a/\lambda$ of the inter-ion spacing $a$ to screening length $\lambda$.  At $0.010\unit{fm}^{-3}$ the system has an average of $\langle{Z}\rangle\simeq78$, temperature of $1\unit{MeV}$ and inter-ion spacing $a=(3/4\pi n_{\text{ion}})\simeq16.7\unit{fm}$.  The ion density $n_{\text{ion}}$ is obtained from $n_{\text{ion}}=Y_p n/\langle{Z}\rangle$.
Thus, $\kappa\simeq1.67$ and the Coulomb parameter $\Gamma$ is
\begin{equation}
 \Gamma=\frac{\langle{Z}\rangle^2e^2}{aT}\simeq530.
\end{equation}

According to Fig. 1 of Ref. \cite{PhysRevE.66.016404} a system with these parameters should form a $bcc$ lattice, provided we ignore the free energy of mixing of ions of different charges \cite{PhysRevE.81.036107}.  
We note here that the definitions of $\kappa$ and $\lambda$ in Ref.  \cite{PhysRevE.66.016404} are slightly different from ours.  
Vaulina \textit{et al} define $a=n^{-1/3}$ so our value of $\kappa$ ($\Gamma$) should be multiplied (divided) by a factor of $(3/4\pi)^{1/3}$ in order to be compared to theirs.  
This, however, does not change our conclusion.

\begin{figure}[h]
\centering
\includegraphics[width=0.5\textwidth]{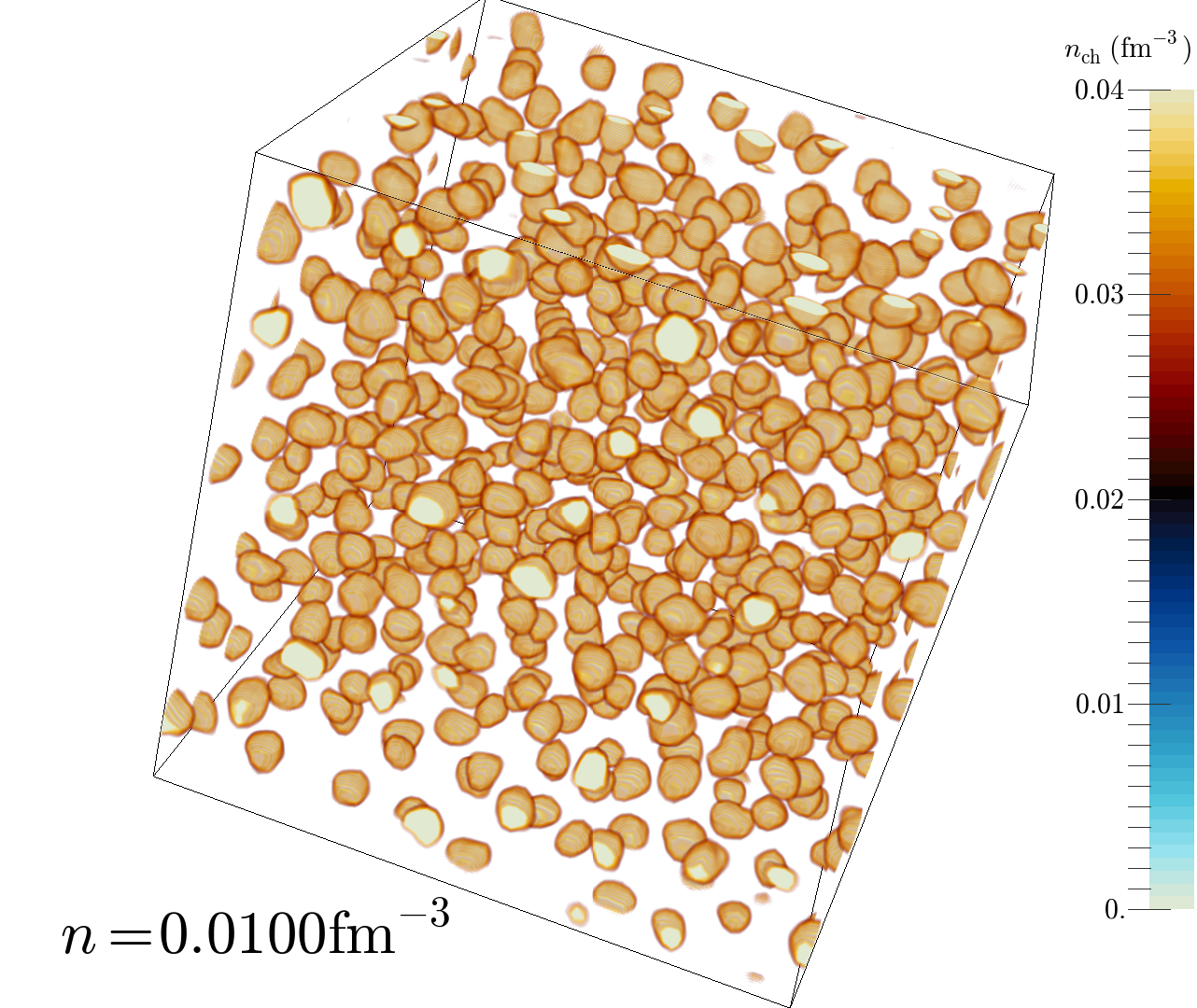}
\caption{\label{fig:random} System at a density of $0.010\unit{fm}^{-3}$ obtained from an initial random configuration at the same density. This figure was generated using ParaView \cite{Paraview}.}
\end{figure}

The system obtained from a random start and evolved at constant density also has very few small nuclei $A\lesssim10$, while most nuclei are in the range $A\sim40$ to $150$, see bottom plot of Fig. \ref{fig:hist}.
The nuclei in this sytem do not form a lattice structure, but rather form a liquid-like structure as its average charge $\langle{Z}\rangle=35.7$ implies a value of $\Gamma\simeq142$ and $\kappa\simeq1.29$.
Clear differences between this system and the one generated from slowly stretching the box can be seen by comparing Figs. \ref{fig:bcc} and \ref{fig:random}.
This system forms smaller nuclei than the slowly stretched system because of the effects of the Coulomb barrier.  From their initial random positions nucleons start to correlate to form nuclei. 
As they group together the charge of the nuclei reaches a value that makes it increasingly difficult for other small nuclei to merge with it.  This may be a disadvantage of treating the system classically because quantum tunneling is neglected.

Unlike the two systems just discussed at $0.010\unit{fm}^{-3}$, the configuration obtained from stretching the box at a rate of $\dot\xi=1.0\times10^{-5}\unit{c/fm}$ has both several small nuclei $A\lesssim20$ and some large elongated nuclei, $A\gtrsim300$ with a peak in mass number of $A\sim150$, see the center plot in Fig. \ref{fig:hist}.  This difference arrises because this system did not have enough time to equilibrate.   The large elongated nuclei have not had time to fission into smaller nuclei.  It is likely that the equilibration time for the pasta structures shown here is much larger than the very roughly $t\approx 100,000\unit{fm/c}$ time scale for significant density changes in a system stretching at a rate of $\dot\xi=1.0\times10^{-5}\unit{c/fm}$.

\subsubsection{Simulation Visualizations}
\label{ssec:movies}
Movies that show charge density isosurfaces versus density for the above simulations are available on line (and we anticipate directly from Physical Review should this article be published).  
Description of the movies and web page address where they are available can be found in Tab. \ref{tab:movies}.

\begin{table*}[t]
\caption{\label{tab:movies} Web page addresses for visualization of simulations described in the text. Each line describes the file name, the stretching rate of the run and its {\tt{URL}} and size in MB. Visualizations files were generated using ParaView Software \cite{Paraview}.
}
\begin{ruledtabular}
\begin{tabular}{c|cc|c}
File       & $\dot\xi\unit{(c/fm)}$ & URL & Size\,(MB) \\
\hline
 \href{https://iu.box.com/s/rusp1yiqpba39rfhgo9m}{51200\_40\_1\_1e-7\_720\_no\_stereo.avi} & $1.0\times10^{-7}$  &  
 \href{https://iu.box.com/s/rusp1yiqpba39rfhgo9m}{\tt{https://iu.box.com/s/rusp1yiqpba39rfhgo9m}} & 90.8 \\
 \href{https://iu.box.com/s/eq29snty9a9w73v8pfr7}{51200\_40\_1\_1e-6\_720\_no\_stereo.avi} & $1.0\times10^{-6}$  &  
 \href{https://iu.box.com/s/eq29snty9a9w73v8pfr7}{\tt{https://iu.box.com/s/eq29snty9a9w73v8pfr7}} & 90.8 \\
 \href{https://iu.box.com/s/9tgme4ti2h9y35h3876p}{51200\_40\_1\_1e-5\_720\_no\_stereo.avi} & $1.0\times10^{-5}$  &  
 \href{https://iu.box.com/s/9tgme4ti2h9y35h3876p}{\tt{https://iu.box.com/s/9tgme4ti2h9y35h3876p}} & 45.8 \\
 \href{https://iu.box.com/s/f6ld58n8z0336ix8e53n}{51200\_40\_1\_1e-6\_720\_no\_coulomb.avi} & $1.0\times10^{-6}$\footnote{Does not include the Coulomb potential.} &  
 \href{https://iu.box.com/s/f6ld58n8z0336ix8e53n}{\tt{https://iu.box.com/s/f6ld58n8z0336ix8e53n}} & 29.8 \\
\end{tabular}
\end{ruledtabular}
\end{table*}

\subsubsection{Minkowski Functionals}\label{ssec:MF}

\begin{figure}[h]
\centering
\includegraphics[width=0.5\textwidth]{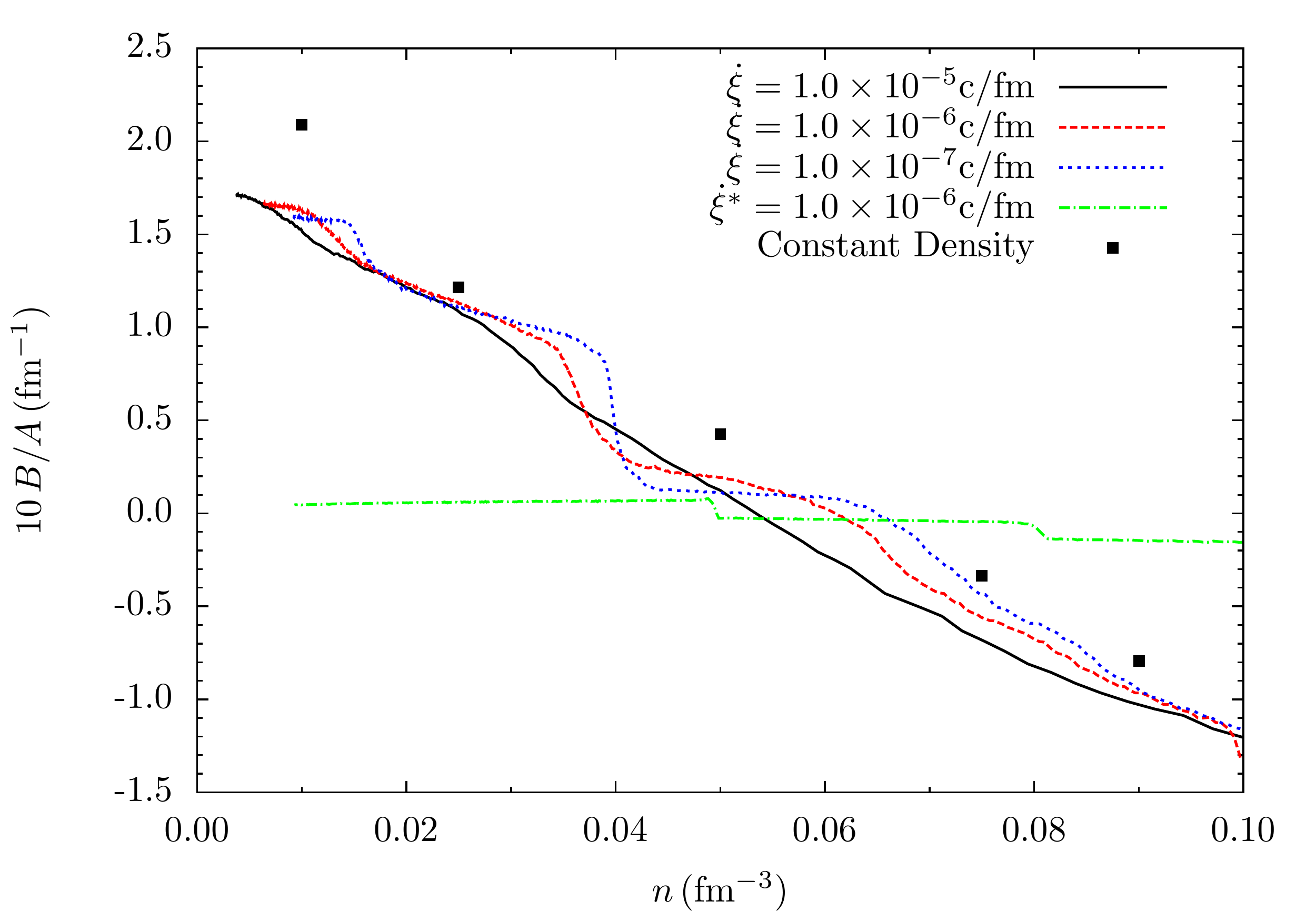}
\caption{\label{fig:breadth} (Color online) Normalized mean breadth $B/A$ as a function of the density $n$ for three full calculation using different stretch rates and one calculation ignoring Coulomb interactions. Results are compared to five computations done at constant densities of $0.010$, $0.025$, $0.050$, $0.075$ and $0.090\unit{fm}^{-3}$.}
\end{figure}

\begin{figure}[h]
\centering
\includegraphics[width=0.5\textwidth]{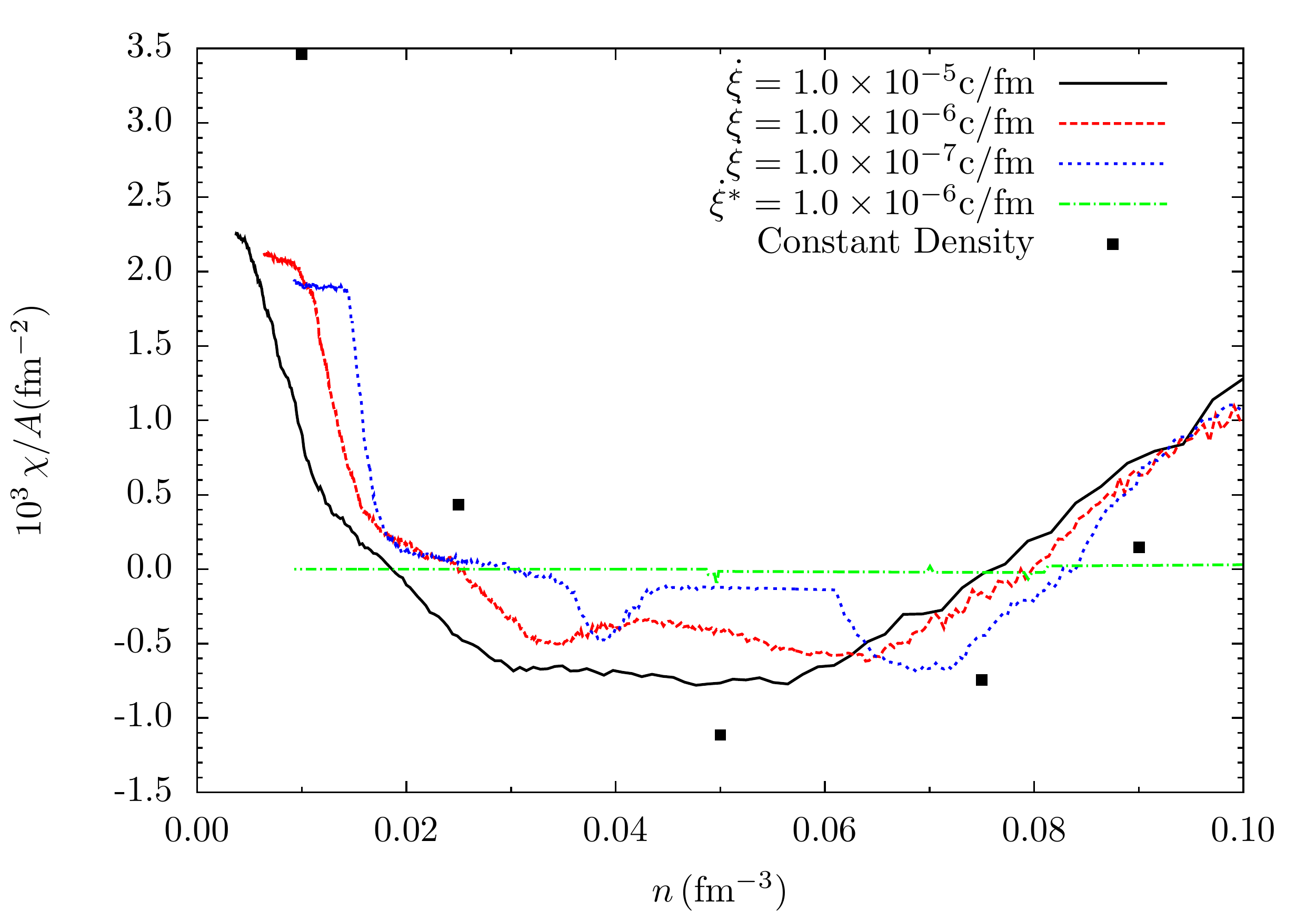}
\caption{\label{fig:chi} (Color online) Lines represent the normalized Euler characteristic $\chi/A$ as a function of the density $n$ for three full calculation using different stretch rates and one calculation ignoring Coulomb interactions. Results are compared to five computations done at constant densities of $0.010$, $0.025$, $0.050$, $0.075$ and $0.090\unit{fm}^{-3}$.}
\end{figure}

The evolution of the normalized mean breadth $B/A$ and Euler characteristic $\chi/A$ as a function of the density $n$ for three runs done using different stretching rates, $\dot\xi=1.0\times10^{-5}$, $1.0\times10^{-6}$ and $1.0\times10^{-7}\unit{c/fm}$, are shown in Figs. \ref{fig:breadth} and \ref{fig:chi}, respectively.  Also shown in these figures are the results obtained for the constant density runs at $0.010$, $0.025$, $0.050$, $0.075$ and $0.090\unit{fm}^{-3}$ and for a stretching run with $\dot\xi=1.0\times 10^{-6}$ and no Coulomb interactions.

First we note that by comparing the systems that include and do not include the Coulomb potential, it becomes clear that it is the competition between nuclear and Coulomb forces that gives rise to the richness of the pasta shapes.  Without this competition the shapes accessible to the pasta phase are very limited, as noted by the small values for the normalized Minkowski functionals for the simulation that does not include the Coulomb potential.


We also note that systems stretched at rates $\dot\xi$ of $1.0\times10^{-5}$ and $1.0\times10^{-6}\unit{c/fm}$ do not have enough time to equilibrate so that the transitions between shapes are not as sharp as in the system stretched at $\dot\xi=1.0\times10^{-7}\unit{c/fm}$.
If the system is stretched slow enough, $\dot\xi=1.0\times10^{-7}\unit{c/fm}$, the pasta shapes have more time to equilibrate and the transitions between shapes are sharper.

In the plot shown in Fig. \ref{fig:breadth} for the normalized mean breadth $B/A$ we see that the faster a system is stretched the smoother is the mean breadth dependence on the density $n$.
For a stretch rate of $\dot\xi=1.0\times10^{-5}\unit{c/fm}$, $B/A$ increases almost linearly as the density decreases.  If the expansion rate is decreased to $\dot\xi=1.0\times10^{-6}\unit{c/fm}$ some kinks in the curve of $B/A$ appear.  The two most prominent ones are at $0.040\unit{fm}^{-3}$, the transition region from the lasagna to the spaghetti phase and $0.015\unit{fm}^{-3}$, the transition region from the spaghetti to the gnocchi phase.

When the expansion rate is decreased further, $\dot\xi=1.0\times10^{-7}\unit{c/fm}$, the kinks in the curve of $B/A$ become even larger and are displaced to slightly higher densities.  The mean breadth $B$ in the region from $0.040\unit{fm}^{-3}$ to $0.060\unit{fm}^{-3}$ is close to zero indicating that the pasta shapes have almost zero curvature in all directions.  This is characteristic of the lasagna phase as it is formed of nearly flat sheets of nuclear matter.   At a density of $0.040\unit{fm}^{-3}$ the transition region from the lasagna to the spaghetti phase becomes even sharper than before.
Also the transition from the spaghetti to the gnocchi phase happens at a higher density, $0.015\unit{fm}^{-3}$, than for the system stretched at $\dot\xi=1.0\times10^{-6}\unit{c/fm}$, $0.012\unit{fm}^{-3}$.

Figure \ref{fig:chi} shows the normalized Euler characteristic $\chi/A$ as a function of density $n$. 
We note again that the curves become smoother as the expansion rate increases showing the system does not have time to reach equilibrium when stretched too fast.  In this plot the transitions from the spaghetti to the gnocchi phase becomes clearer for the slowest expansion rate.
Another important thing to note is that, when looking at both plots, it is clear that for lower densities the topology of the systems obtained from expansion runs do not converge to the values obtained from the constant-density runs.  This suggests that the constant density runs (al least) have not equilibrated.

\section{Conclusions}\label{sec:Conclusions}

In this paper we studied the dynamics of pasta phase transitions using an MD formalism to isothermally expand nuclear matter from densities of $0.100\unit{fm}^{-3}$ to $0.010\unit{fm}^{-3}$ or less.   Expansions were performed at a temperature of $T=1\unit{MeV}$ and proton fraction of $Y_p=0.40$ using stretching rates of $\dot\xi=10^{-5}$, $10^{-6}$ and $10^{-7}\unit{c/fm}$, see Eq. \ref{eq:stretch}.  These runs were then compared to constant-density runs at densities of $0.010$, $0.025$, $0.050$, $0.075$ and $0.090\unit{fm}^{-3}$.

For each run we obtained the Minkowski functionals as a function of density and learned that, for MD simulations such as ours, the methods used to prepare the pasta alter the resulting pasta shapes significantly.  First, we noticed that the Coulomb force is essential for the formation of pasta-like configurations.   Without Coulomb, the pasta shapes are confined to  small values of both the mean breadth  $B$ and Euler characteristic $\chi$.

When the Coulomb potential is included, we obtain pasta configurations similar to the ones obtained by other works using several other methods as discussed in Sec. \ref{sec:Intro}.
These shapes are spherical-holes, cylindrical-holes, sheets (lasagna), cylinders (spaghetti) and spheres (gnocchi).  However, the geometrical shapes accessible to the pasta phases in our simulations as well as their arrangement were dependent on the expansion rates.
In the simulation with the slowest expansion rate, $\dot\xi=10^{-7}\unit{c/fm}$, close to periodic structures are formed in the gnocchi, spaghetti and lasagna phases in addition to the periodicity imposed by the periodic boundary conditions.  Meanwhile the pasta shapes obtained from fast expansion rates, $\dot\xi=10^{-5}$ and $10^{-6}\unit{c/fm}$, did not exhibit any additional translational symmetry beyond the enforced ones.  We recall here that calculations that used volumes larger than the Wigner-Seitz cell obtained these additional translational symmetries for the pasta shapes \cite{Okamoto2012284,PhysRevC.69.055805,PhysRevC.77.035806,PhysRevLett.94.031101,PhysRevLett.103.121101} which had not been observed by the larger MD calculations of Horowitz \etal \cite{PhysRevC.69.045804,PhysRevC.70.065806,PhysRevC.72.035801,PhysRevC.78.035806}.

Another point to note is that the two simulations with fastest expansion rates have not reached equilibrium.  This can be argued from the fact that the faster the expansion rate the longer, as a fraction of total run time, different pasta phases coexisted.  However, for the slow expansion rate, the coexistence of pasta shapes did not last as long, as a fraction of total run time.  Thus, for fixed temperature and proton fraction, as the expansion rate gets closer to being quasi-static the transition between phases become more abrupt.  This suggests that the transition between different pasta phases is first-order.  The plots of $B/A$ and $\chi/A$ for the slowest expansion run show that the transitions between lasagna to spaghetti and spaghetti to gnocchi phases are much sharper than in the fast expansion runs.

The sudden changes in pasta shapes, $B/A$ and $\chi/A$, as the density of the system is decreased can help us place lower limits on the density where each pasta phase occurs.
Because of the way the system is evolved a pasta phase may be kept in a metastable state at a lower density than it would normally exist.
By this reasoning we claim that for a proton fraction of $Y_p=0.40$ and temperature $T=1\unit{MeV}$ the lasagna phase occurs for densities of $n\simeq0.040-0.060\unit{fm}^{-3}$.
The spaghetti phase occurs for densities of $n\simeq0.018-0.40\unit{fm}^{-3}$.  Finally the gnocchi phase occurrs below a density of $0.018\unit{fm}^{-3}$.

These simulations also set a time scale for the transition between the lasagna/spaghetti phases and the spaghetti/gnocchi phases.  
The transition from gnocchi to spaghetti phase happens between $n\simeq0.0145-0.018\unit{fm}^{-3}$ and takes a time of approximately $1\,300\,000\unit{fm/c}$.
Meanwhile, the transition from spaghetti to lasagna phase happens between $n\simeq0.038-0.044\unit{fm}^{-3}$, and takes a time of approximately $600\,000\unit{fm/c}$.

In Ref. \cite{PhysRevLett.103.121101} Watanabe \etal obtained the transition time using QMD from adiabatic compression runs starting at a temperature of $T=0.25\unit{MeV}$ for the transition between the spherical (gnocchi) and cylindrical (spaghetti) phases for a similar proton fraction, $Y_p=0.39$.  In their work, the transition between these two phases happens between $n\simeq0.040-0.045\unit{fm}^{-3}$ and the transition time was of the order of $10\,000\unit{fm/c}$.
This much shorter transition time may be because of the momentum dependent QMD interactions that may increase the nucleon momenta in comparison to our momentum independent interactions.  
Furthermore some of the difference in transition time may arise because the QMD simulations were for significantly smaller systems using fewer nucleons than the 51\,200 that we use.

We also observed that constant-density runs with nucleons initially assigned random positions exhibit some pasta-like shapes such as cylindrical-holes, cylinders and spheres.  
These runs, however, where not able to produce spherical-holes and flat sheets as the expansion-runs did.
It is likely that the spherical-hole phase appears at higher density for this type of simulation while the lasagna phase may take a very long time to form from the initial conditions chosen.
Also, for the low density run $n=0.010\unit{fm}^{-3}$, the Coulomb barrier and the classical character of our simulation prevented the formation of large nuclei.

Our simulations explicitly demonstrate nucleation mechanisms for each of the observed pasta phase transitions.  First spherical holes were observed to nucleate from density fluctuations in an originally uniform system.  This is very similar to the nucleation of vapor bubbles for a conventional liquid-gas phase transition.  However, the Coulomb interaction keeps the spherical holes from growing to very large sizes, as occurred in our simulation with out Coulomb interactions.

These spherical holes were observed to merge, with a further decrease in density, to form cylindrical holes (``anti-spaghetti'').  Next the anti-spaghetti became cross linked and finally quickly merged to form the lasagna phase.  As the density was decreased still further, holes appeared in the lasagna planes and these holes grew to convert the lasagna into a cross linked network of spaghetti.   These holes, in the cross linked network of spaghetti, lead to a negative excursion in $\chi/A$ shown in Fig. \ref{fig:chi} near $n=0.04\unit{fm}^{-3}$ for the run with $\dot\xi=10^{-7}\unit{c/fm}$.
The cross links disappeared at lower density to produce isolated nearly straight spaghetti strands.  Finally these spaghetti strands rapidly fissioned to form nearly spherical nuclei.  

If the transition, with changing density, from spaghetti to spherical nuclei is reversible, then the criteria for the first formation of non-spherical pasta phases is related to when the spaghetti strands become unstable to fission.  This depends on the sizes of the coulomb and surface energies. Thus pasta formation is related to nuclear fission.       

In future work we will perform simulations with even slower expansion rates starting above saturation density to further determine when the pasta is equilibrated.
Once the pasta is equilibrated we may study the observed low energy bending modes of the spaghetti and lasagna shapes that may contribute significantly to the heat capacity even at low temperatures.
In addition, we will explore finite size effects by performing larger simulations with more than 51200 nucleons. 
Finally, we will perform simulations with a range of smaller proton fractions. 
These simulations will then be used to calculate a variety of observable.
Neutrino opacities can be determined using the formalism in Refs. \cite{PhysRevC.69.045804,PhysRevC.70.065806} 
while the bulk viscosity may be obtained by homogeneous periodic compressions and expansions of the system \cite{PhysRevA.21.1756}.
We also intend to determine the shear modulus and breaking strain of the pasta phases by observing the response of the system to deformations of the simulation volume.

\begin{acknowledgments}

We are grateful to David Reagan at the Advanced Visualization Laboratory - Indiana University for his help with ParaView.  We would also like to thank Indiana University for access to the BigRed supercomputer.  This research was supported in part by DOE grants DE-FG02-87ER40365 (Indiana University) and DE-SC0008808 (NUCLEI SciDAC Collaboration) and by the National Science Foundation through XSEDE resources provided by the National Institute for Computational Sciences under grant TG-AST100014.

\end{acknowledgments}

\bibliography{Pasta5}

\end{document}